\documentclass[12pt,preprint]{aastex}

\slugcomment{Accepted by the Astrophysical Journal}

\shorttitle{Quasar Radio Variability}
\shortauthors{Barvainis et al.}

\begin{document}

\title{Radio Variability of Radio Quiet and Radio Loud Quasars}

\author{Richard Barvainis\footnote{Any opinions, findings, and conclusions 
and recommendations expressed in this material are those of the author and 
do not necessarily reflect the views of the National Science Foundation. }}
\affil{National Science Foundation, 4201 Wilson Blvd, Arlington, VA 22230 USA;}
\affil{and Department of Physics, Gettysburg College, Gettysburg, PA 17325} 
\email{rbarvai@nsf.gov}

\author{Joseph Leh\'ar}
\affil{CombinatoRx, Inc, 650 Albany Street Boston, MA 02118, USA}
\email{lehar@rcn.com }

\author{Mark Birkinshaw}
\affil{Dept. of Physics, University of Bristol, Tyndall Avenue, Bristol, BS8 1TL UK}
\email{mark.birkinshaw@bristol.ac.uk }

\author{Heino Falke}
\affil{Max-Planck-Institut f\"ur Radioastronome, Auf dem H\"ugel 69, 53121 Bonn, Germany}
\email{hfalcke@mpifr-bonn.mpg.de }

\author{Katherine M. Blundell}
\affil{Oxford University Astrophysics, Keble Rd, Oxford, OX1 3RH, UK}
\email{kmb@astro.ox.ac.uk }
\begin{abstract}
The majority of quasars are weak in their radio emission, with flux densities comparable 
to those in the optical, and energies far lower.  A small fraction, about 10\%, are hundreds 
to thousands of times stronger in the radio.   Conventional wisdom holds that there are two classes 
of quasars, the radio quiets and radio louds, with a deficit of sources 
having intermediate power.  Are there really two separate populations, and if so, is the 
physics of the radio emission fundamentally different between them?  
This paper addresses the second question, through a study of radio variability across the 
full range of radio power, from quiet to loud.   The VLA was used during 10 
epochs to study three carefully selected samples of 11 radio quiet quasars, 
11 radio intermediate quasars, and 8 radio loud quasars.  A fourth sample consists of 
20 VLA calibrators used for phase correction during the observations, 
all of which are radio loud.  The basic findings are that the root mean square 
amplitude of variability is independent of radio luminosity or radio-to-optical 
flux density ratio, and that fractionally large variations can occur on timescales of months
or less in both radio quiet and radio loud quasars.   Combining this with similarities 
in other indicators, such as radio 
spectral index and the presence of VLBI-scale components, leads to the suggestion 
that the physics of radio emission in the inner regions of all quasars is essentially 
the same, involving a compact, partially opaque core together 
with a beamed jet.   It is possible that differences in large scale radio structures between
radio loud and radio quiet quasars could
stem from disruption of the jets in low power sources before they can escape their host
galaxies.    

\end{abstract}

\keywords{Quasars; Active Galaxies; Radio Emission} 

\section{Introduction}

Historically, quasars have been thought to come in two flavors
in terms of their radio emission, radio loud and radio quiet. The
term ``quiet'' in this context means weak, rather than silent.
Evidence for bimodality in radio strength has been presented by
Kellermann et al.\ (1989) in a VLA study of optically-selected
Palomar-Green Bright Quasar
Survey (BQS) objects, and by others (e.g., Strittmatter et al.\ 1980; Stocke et
al.\ 1992). Radio ``loudness'' is often parameterized by $R$, the ratio of
centimeter radio to optical flux densities, and a histogram of this parameter
plotted by Kellermann et al.\ (1989) appears to show two peaks, a strong
one at $R \lesssim 0.1 - 3$ and another weaker peak at $R \sim 100 -
1000$, with a dip in source number in between.  These two peaks define what
are commonly thought of as the radio quiet (RQQ) and radio loud (RLQ) quasar
classes, with the large majority of quasars ($\sim 90\%$) belonging to
the radio quiet catagory.  The sources in between ($3<R<100$) have more
recently come to be called radio intermediate quasars, or RIQs.  Miller,
Rawlings, \& Saunders (1993), and  Falke, Sherwood, \& Patnaik (1996) suggested
that RIQs could be the relativistically boosted counterparts of RQQs.  
The break point in radio luminosity between loud and quiet quasars is generally
taken to be $L_r = 10^{25}$ W Hz$^{-1}$ (e.g., Hooper et al.\ 1995; 
Miller, Peacock, \& Mead 1990).

   However, the Bright Quasar Survey, used by both Kellermann et al.\ (1989) and
Stocke et al.\ (1992), has been claimed to be only $\sim 30\%$ complete
(Goldschmidt et al.\ 1992).   
More recent studies based on large, carefully 
chosen complete samples have found mixed evidence for bimodality 
in radio power. 
Hooper et al.\ (1995) performed a statistical analysis of radio and 
optical measurements of 256 optically-selected quasars from the 
Large Bright Quasar Survey (LBQS),  
and found the results ``ambiguous" (their term) regarding
bimodality.  Using similar methods applied to 87 optically-selected quasars 
from the Edinburgh Quasar Survey, Goldschmidt et al.\ (1999) reached a 
similar conclusion.
No hint of bimodality is seen in the First Bright Quasar Survey (FBQS), 
a {\it radio-selected} quasar sample that shows a continuous distribution of 
radio flux densities (White et al.\ 2000; Becker et al.\ 2001), but this
may be biased toward the radio intermediate and radio loud populations.
Indeed, this was argued to be a selection effect by Ivezi\'c et al.\ (2002), who 
found a bimodal distribution of $R$ in a large sample of SDSS quasars. Cirasuolo
et al.\ (2003) challenged Ivezi\'c et al.\ on statistical grounds, and 
the debate goes on.  

   A pertinent question then is whether there are fundamental differences
between RQQs, RIQs, and RLQs, other than the magnitude of radio core power and 
differences in associated large scale structure.  Large-scale radio 
structures fall into two classes (Fanaroff \& Riley 1974), FR I (brightness 
decreasing with distance from core) for low
power radio sources, and FR II (edge-brightened, classical doubles) for
high power radio sources.  RQQs and RLQs are 
extremely similar in both continuum and spectral line properties across 
the rest of the electromagnetic
spectrum, except in the case of blazars, where
the synchrotron and self-Compton emission from the compact radio core is so
strong that it dominates other components.  Even in the radio, RQQs and RLQs 
have similar core spectral index distributions (Barvainis, Lonsdale,
\& Antonucci 1996), and similar parsec-scale radio structures 
(compact cores and jets; Blundell \& Beasley 1998; Ulvestad, Antonucci, 
\& Barvainis 2004).  It also appears, based on deep radio
imaging by Blundell \& Rawlings (2001) of the borderline RQQ/RIQ E1821+643,
that RQQs can have extended radio sources much like those seen
in low-power radio galaxies (FR I objects).

Until recently, quasar environments and host galaxy morphologies
seemed to differentiate quasar radio types.
For nearby, low luminosity AGNs,
radio loud objects reside almost exclusively in elliptical hosts, while
radio quiets are found most often in spirals.  However, at higher
luminosities, above the traditional $M_B = -23$~mag dividing line between Seyfert
galaxies and quasars, a study by Dunlop et al.\ (2003) 
using Hubble Space Telescope imaging concluded
that true quasars reside in ellipticals regardless of
radio power.  Furthermore, the galaxy
environments of RQQs and RLQs appear to be indistiguishable in richness, 
according to other recent work (McLure \& Dunlop 2001; Wold et al.\ 2001).

So the old problem of the physical distinctions between RQQs and RLQs is
even more relevant today.  Are there differences  
other than radio strength and large-scale morphology?  
One area yet to be carefully examined
in this context is radio variability.  In the past, emphasis has been 
placed on studying the variability of radio loud active galactic 
nuclei (AGNs), especially blazars (e.g., Wagner \& Witzel 1995; Aller et al.\ 1999). 
In these objects the radio emission
is certainly due to a relativistically beamed jet observed end-on, and 
one goal of 
multi-wavelength monitoring is to understand particle 
acceleration processes in the jet plasma as well as the relativistic 
effects associated with the changing geometry and structure of jets.

On the other hand, for radio weak AGNs -- here meant to include RQQs and
RIQs but not RLQs -- the 
database is much more sparse.  Few surveys exist that address 
the issue
of radio variability in either RQQs or low-luminosity AGNs such as Liners 
and dwarf-Seyferts (for preliminary studies see Barvainis, Lonsdale \& 
Antonucci 1996, and Ho et al.\ 1999).
In many of these cases we are not even entirely sure that the radio 
emission is related to the AGN itself, rather than, for example, 
a starburst.

Clearly, if (some of) the radio emission in radio-weak AGNs is coming from 
the central engine, 
we would expect to see a certain degree of radio variability.  The 
amount of varibility at different radio powers should provide 
insight into the degree to which the physics of radio emission is similar, 
or different, in
radio loud, intermediate, and quiet quasars.  With these issues as motivation, 
we have carried out a campaign using the VLA to monitor radio variability in 
50 quasars spanning a range of
flux density densities from sub-milliJansky to over a Jansky.  This paper reports
the results of this campaign and discusses implications for the radio 
loud/quiet quasar phenomenon.

\section{Source Selection and Observations}

Quasar variability could in principle be a function of optical luminosity,
radio/optical ratio $R$, or redshift, so we selected targets that cover 
a wide range of all three parameters, but which fall within a right
ascension range compatible with our initial VLA time allocation.  We 
selected all of the PG (Kellermann et al.\ 1989) and LBQS 
(Visnovsky et al.\ 1992; Hooper et al.\ 1996; Hooper et al.\ 1995)
quasars which had a flux density  
$> 0.3$ mJy at 8.6 GHz.  We then removed all lobe-dominated 
targets and those which had already been extensively monitored. 
Due to a dearth of intermediate-redshift targets, we arbitrarily
selected seven more radio quiet targets from those $\sim 30$ Veron-Cetty 
quasars in our target sky region whose radio flux densities had been 
detected using the NVSS (see Bischof \& Becker 1997).  Table 1a summarizes 
the properties of this original set of targets.  During the VLA observations a number
of phase calibrators were also observed following standard procedures, and the 
variability properties of these sources can also be investigated
using the data thus obtained.  The calibrator source list can be found in 
Table 1b.  

The breakdown in source numbers between the various classes is 11 RQQs, 11 
RIQs, and 8 RLQs among the program sources, and 20 calibrators (all RLQs).

A log of the observations is given in Table 2, where it can 
be seen that a variety of VLA configurations were used over 
10 epochs at X-band (8.48 GHz).  In the final run the sources
were  also observed
at L- and C-bands (1.49 and 4.89 GHz) to gather spectral information.  
The first nine monitoring epochs had spacings from about 2 weeks to 
2 months, due in part to the difficulties inherent in 
regular VLA scheduling and in part to the wish to obtain a variety
of time-samplings.  The tenth epoch, intended to provide 
an extended time baseline, occurred almost 19 months after the ninth.
The initial
epoch had the longest VLA baselines, and to reduce phase noise and
better match the angular resolution of later epochs the UV data
were tapered by 160 k$\lambda$.  This yielded a beam size of 
$\sim 1.3'' \times 1''$, which is similar to the B-array beam
size.  Beam sizes for the different epochs/configurations 
are listed in Table 2.

Observations were carried out in the standard manner, using one nearby 
phase calibrator for each target source.  Integration times ranged 
between 2--12 minutes depending on the 8.6 GHz flux density of each target.

Absolute flux density calibration was done against 3C48.  When the light curves 
were examined it became clear that there was some repeating
structure in the target and phase calibrator light curves 
(many light curves for weakly variable
sources had a similar shape) due to unknown causes.  One 
possibility is effects related to different 
array configuations and beam sizes on 3C48, which is not a point source.
This repeating structure has been corrected using one of the 
phase calibrators, 2210+016, which is a compact steep spectrum (CSS) object 
and therefore less likely than other sources to vary or have its observed 
flux density affected by VLA configuration.  The correction
factors were in the range $\pm 5$\%.  This procedure appears to have removed
the systematic structure effectively, and we think the calibration
is now accurate. See note on error bars below.

All sources in the original sample were point-like on the maps except 
P0003+158, P2251+11, and P2308+098, 
which are triples (core plus double lobes).  We dropped P2251+11 because
of possible confusion of core and lobe flux density (only the cores are of interest
for this project), and because we did not
have a spectrum for the core alone.  P0003+158 and P2308+098 were 
both separable between core and lobes at X-band and we
believe the core flux densities are accurate, except in D-array (final epoch) for 
P0003+158.  This last point for this source 
was dropped from the tables and figures.  To construct a spectrum for P0003+158 
we used higher-resolution data from the literature (Hutchings et al.\
1996).  The phase calibrator 2210+016 is not reported 
in the tables, because it was used as a secondary flux density calibrator and 
thus by definition did not vary (see discussion above).  Because we
concentrated on unresolved cores in this project, we believe that 
the changing VLA configurations/resolutions at different epochs is not a 
significant cause of flux density variations in the target sources.  There
is no evidence that the flux densities are systematically higher in the more
compact configurations as would be expected if extended flux density were 
being picked up.

\section{Data Reduction and Analysis}

\subsection{Maps and Fluxes}

Maps were made using AIPS for all target sources.  For
sources with flux
densities above 15 mJy phase self-calibration was performed, which
reduces the effects of phase noise on the flux density measurements.  
In sources too weak for self-calibration, phase noise is a potential issue
in obtaining accurate flux densities, since it can remove flux density from 
the central beam and spread it throughout the map.   Care was taken to examine 
stronger target sources and calibrators to ensure that phase noise was not
a significant problem during the observations of the weak sources.  However, 
we estimate phase noise errors up to 6\% in some cases, and 
conservatively use this figure as the calibration uncertainty on 
all sources which were not self-calibrated.  The calibration 
uncertainty adopted for strong ($> 15$ mJy) sources is 2\%.   Data for 
all sources below 15 mJy was discarded for one observing 
epoch (4/3/97) because of bad weather and resultant
high phase noise for the whole run. 

  Flux densities for the target sources and the calibrators 
are reported in Tables 3a and 3b, and represent the peak flux density at the
source location on the map for program sources.   Maps were not made for
the calibrators; rather, the reported flux density is from a fit performed in the 
UV plane as part of the standard calibrator reduction process (AIPS task GETJY)
that includes self-calibration.

\subsection{Debiased Variability Index} 

The light curve for any source with error bars on the measured flux densities 
will show some variation, even if the source itself is constant.  To develop a 
quantitative measure of source 
variations we use the debiased RMS, which corrects for uncertainties in 
the flux densities.  The root mean square variation, after debiasing, is then 
used to compare one source to another, or one group of sources to another.
The debiased variability index is defined here as a percentage of 
the mean flux density, 

\begin{equation}
V_{RMS} = 100\cdot{\sqrt{\Sigma\left(S_\nu(t)-\left<S_\nu\right>\right)^2 -
\Sigma\sigma^2}\over N\cdot\left<S_\nu\right>},\end{equation}
where $N$ is the number of data points, $\sigma$ is the measurement error,
and $\left<S_\nu\right>$ is the mean flux density 
(see www.stat.psu.edu/$\sim$mga/scca/q\_and\_a/qa25.ps;
Akritas \& Bershady 1996). We set the index to be negative when
the value inside the square root becomes negative (i.e., for
non-variable sources where the error bars are too conservative, or in cases where
random chance causes the measured variations to be smaller than expected from
the error bars).

The debiased RMS for each source and calibrator is given in Tables 1a and
1b.  Also reported are the redshift, the optical V magnitude, the mean flux density
over the course of the program, the radio-to-optical ratio $R$, and 
the radio spectral index between C-band (4.89 GHz) and X-band (8.48 GHz).  
We define spectral index $\alpha$ by way of $S_{\nu} \propto \nu^{\alpha}$.

Eye estimates of the maximum fractional change in flux density of
the variable sources were also made, and are shown as $f_{\rm var}$ in
Tables~1a and~1b with their associated observed timescales $t_{\rm
var}$. $f_{\rm var}$ and $t_{\rm var}$ are not quoted with errors
in the Tables because the relatively short span of these
observations makes it impossible to define whether a particular
change of flux density is typical of the overall behaviour of that
source, or even to define the mean or quiescent level of the source
(if such concepts have a well-defined meaning). However, they do
give a qualitative indication of the apparent properties of the variable
sources, in the sense that large $f_{\rm var}$ and small
$t_{\rm var}$ sources are those with the largest and fastest pattern
of variability in these data.

\section{Results}

Figure 1 shows the the light curves for the program sources.  The radio 
spectrum is shown as a shaded inset box for each source.  The 
calibrator light curves are shown in Figure 2.

Several of the sources exhibit large-amplitude variations in brightness
within the 1997 dataset -- L0004+0224 being an RQQ example, and
P2209+184 being an RLQ example. In each case
substantial fractional variations in brightness were seen on timescales of
less than three months. Taking account only of the change in
flux density, and the corresponding timescales, we calculate that the
brightness temperatures of the emission from the RQQ L0004+0224 and
the RLQ P2209+184 are $3.3 \times 10^{10}$ K and $1.1 \times 10^{11} \
\rm K$, respectively, if no account is taken of the (likely) effects
of relativistic beaming.  These brightness temperatures are
at or just below the ``maximum brightness temperature'' of 
$\sim 10^{11}$ K according to the model of Readhead (1994), and it
is interesting that the brightness temperatures for an RQQ and an RLQ 
are comparable, suggesting a
substantial similarity in the radiation environments in these AGN.

Figures 3-7 plot the fractional debiased RMS, $V_{RMS}$, versus various 
quantities: the 
radio/optical ratio $R$, the radio and optical luminosities\footnote{For 
simplicity, luminosities $L_{\nu}$ are calculated assuming 
$\Omega_{\rm matter} = 0.2$ and $\Omega_{\rm lambda} = 0$.  This gives
results within $\sim20\%$ out to $z = 3.5$ relative to $\Omega_{\rm matter} = 
0.3$ and $\Omega_{\rm lambda} = 0.7$.}, the redshift, and the radio 
spectral index.  

There are no strong trends visible in any of the plots, particularly when
P0007+106 (III Zw 2), one of the most highly variable radio sources 
known (see Falcke et al.\ 1999), is ignored.
In Figure 7 there is a weak tendency for the debiased RMS to be 
larger for larger
values of the spectral index (i.e., sources with flat or inverted spectra 
seem to be more variable, on average, than steep-spectrum sources).  
This result is expected, and indeed it is surprising that the effect is not 
stronger.  Removal of
2-3 of the most highly variable sources would erase the trend.
There are many inverted-spectrum sources that did not vary 
or were only weakly variable during the course of this program; these objects 
may have been in temporarily quiescent phases, or may be part of the 
gigahertz-peaked spectrum population, which is known to be only weakly variable.  

Cursory inspection of Figure 3 gives the impression that the calibrators are
more variable than other sources.  However, this is not born out by 
statistics, as presented in Table 4.  There we use the quantity 
$V^*_{RMS}$, which is 
just the debiased RMS with all negative values (meaning that those sources are 
non-variable) set to zero.  The RQQs, RIQs, RLQs, and CALs all have
mean $V^*_{RMS}$ values between 5\% and  8\%.  Given the large dispersions
within each class, any small differences (at the couple of percent level) 
are clearly not statistically significant.

The calibrators might be expected 
to be more variable than the target sources, since the ideal calibrator
would be flat-spectrum (typically defined as $\alpha > -0.5$), compact, 
and core-dominated.  These are all attributes of the 
class of blazars, so-named because of their high radio and optical 
variablity.  So it comes as somewhat of a surprise that
the calibrators are indistinguishable from the target sources, which
were not selected for any of these properties.  
Clearly, ``VLA calibrator" is not equivalent to ``blazar".

Figures 8 and 9 compare the variability and spectral index between 
the four source classes (RQQs, RIQs, RLQs, and CALs) by means of 
histograms.  Within the limitations
of small number statistics and the fairly wide dispersion in values within
each class, especially for spectral index, the distributions do not 
appear to be substantially different.\footnote{We resisted employing statistical comparisons such as the 
Kolmogorov-Smirnov test, which do not take into account error bars and 
for which our sample sizes are considered marginal or too small 
for meaningful results.} 
This is
consistent with the mean and median statistics presented in Table 4.  Note that
the {\it median} spectral index in Table 4 between RQQs and RIQs differs
by $0.39$, versus the difference in {\it mean} spectral index of only $0.02$. 
This is an 
artifact of the gap for RQQs between $-0.5$ and $0.0$, which we suspect is 
again simply a result of small samples.  
There may be a trend for RLQs (including CALs) to have a somewhat flatter
spectral index than RQQs and RIQs.  The RLQs + CALs, on average, are flatter 
than RQQs + RIQs by roughly 0.27 in spectral index, but recall that the 
calibrators are
selected for having flat radio spectra, and the dispersion within all 
classes is 
high (but lowest for the CALs, as might be expected).  These factors make
any possible differences difficult to judge.  

In summary, from Figures 3-9 we see that the general amplitude of quasar radio 
variability does not seem to be a 
function of the radio-to-optical ratio or radio luminosity (``radio loudness"),
optical luminosity, or redshift.  The radio variability amplitude of a quasar may
depend somewhat on radio spectral index, but perhaps less so than might be expected. 

We do not believe that these
data constitute evidence for a strong distinction between the core properties 
of RQQs, RIQs, and RLQs, other than the property they were selected for 
(value of $R$).   In fact, they appear to be remarkably alike.
  
\section{Discussion}

One obvious conclusion to be drawn from these results is that the 
radio emission from radio quiet quasars originates in a compact structure 
intimately associated with the active nucleus.  The alternative hypothesis,
that the emission from radio-weak quasars is from a starburst (e.g., 
Sopp \& Alexander 1991; Terlevich et al.\ 1992), is ruled out.  
Starburst emission comes from regions with size scales of hundreds or 
thousands of parsecs, so substantial fractional variations on the timescales 
seen here are not possible.  Other evidence against starbursts includes
VLBI detections of compact, high brightness temperature cores in several RQQs 
and RIQs (Blundell et al.\ 1996; Falcke, Patnaik, \& Sherwood 1996; 
Blundell \& Beasley 1998; Ulvestad, Antonucci, \& Barvainis 2004), and the 
observation of flat or 
inverted radio spectra in a number of RQQs (Barvainis, Lonsdale, \& 
Antonucci 1996; Barvainis \& Lonsdale, 1997; this paper). 
Kellerman et al.\ (1994) came to a similar conclusion based on the
unresolved cores seen in many RQQs in VLA observations of the PG sample,
suggesting that ``powerful central engines may be as common in the radio
quiet as in the radio loud population."

The evidence presented here suggests that there are few, if any, differences
between the cores of RQQs, RIQs, and RLQs in terms of their variability and 
spectral index 
properties.  While the RQQs/RIQs appear to have a somewhat broader 
spectral index distribution than the RLQs, this is likely an artifact of 
the preponderance of VLA calibrator sources, which were 
selected for having flat(ish) radio spectra, in the RLQ sample.  The presence 
of variability, flat spectra, and compactness in RQQs and RIQs suggest the same 
sort of beaming effects known
to be present in RLQs.  Therefore no distinction can be drawn in the
present study between the
classes in terms of either fundamental radio emission processes or  
orientation effects.  

The RIQ class was first defined by Miller, Rawlings, \& Saunders (1993).  They
hypothesized, based on the ratio of radio to [OIII] luminosity, that RIQs
could be the beamed counterparts of RQQs, i.e., RQQs seen 
down the jet. 
Falke, Sherwood, \& Patnaik (1996) advanced a similar idea after determining
that a large fraction of the RIQs in the Palomar-Green survey are variable and
have flat radio spectra.  We find no evidence here that RIQs are 
beamed and RQQs are not.  Histograms of variability level and spectral index
for the two classes are indistinguishable (given the small number statistics;
see Figures 8 and 9), as are the mean values of these parameters (Table 4). 

As noted earlier, one major difference between RLQs and RQQs is 
the nature of their extended 
radio emission.   RLQs are characterized by powerful, double-lobed, FR II 
radio structures.  Edge-brightened extended lobes do not appear to be present
in RQQs (Kellerman et al.\ 1994), but it is quite possible that RQQs do possess 
extended structures characteristic of FR I radio galaxies (i.e., decreasing 
surface brightness with increasing distance from the core). In a long 
VLA integration, Blundell \& 
Rawlings (2001) detected a 300 kpc FR I radio structure in the 
borderline RQQ/RIQ E1821+643, 
and argue that previous attempts to detect extended emission in RQQs have 
not gone nearly deep enough. Therefore, it appears possible that RQQs and 
RLQs may be analogous in their extended structures to FR I and FR II radio 
galaxies.  Following Blundell and Rawlings (2001), we speculate that the 
basic mechnanism of core radio emission, i.e., a 
synchroton-emitting relativistic jet, is similar in RQQs and RLQs.
In lower power objects the jet may be more likely to disrupt
and dissipate, forming an FR I structure.  In higher power sources the jet
blasts its way relatively unimpeded out of the host galaxy and goes on
to feed extended double lobes.  

The relation between host galaxy type and radio power remains a mystery.  
Could the strong tendency (at least at optical luminosities below the traditional 
Seyfert/quasar border of $M_B = -23$) for radio quiet AGNs to be 
hosted by spirals, and radio louds by ellipticals, be related to central black
hole mass?   This seems unlikely:  Shields et al.\ (2003), in a study of
the black hole-bulge relationship in quasars, find that the black hole
masses of RQQs and RLQs cover the same range, $10^7 - 10^{10}$ M$_{\sun}$.
Confounding the host-galaxy-type distinction is the finding that at higher 
optical luminosities ($M_B < -23.5$) almost all ``true"  quasars, radio 
loud and 
quiet, reside in elliptical hosts (Dunlop et al.\ 2003, Floyd et al.\ 2003).
Indeed, Dunlop et al.\ (2003) claim that their HST imaging results for 
RQQs and RLQs ``exclude galaxy morphology, black hole mass, or black hole 
fueling rate as the primary physical causes of radio loudness." They 
suggest that the answer lies with some other property of the black hole, most
likely spin (e.g., Blandford 2000; Wilson \& Colbert 1995).

One further consequence of the present work is that the variability properties
of RLQs can serve as an effective description of the expected behaviour
for RIQs and RQQs that are being monitored in the radio as probes of 
other phenomena.  For example,
the variability levels we observe can be used to estimate reasonable
magnification ratio limits between the images of gravitationally
lensed RQQs (notably Q2237+030; Falco et al.\ 1996).

\section{Conclusion}

This paper has shown that the basic emission mechanisms in the cores
of radio loud and
radio quiet quasars may be similar, if not identical, except for 
the level of radio power.  
A critical, but challenging, experiment would be to search for 
compact jets exhibiting superluminal motion in RQQs.  There are now, within the 
present dataset,
several RQQs that show high variability and may be good candidates
for observation.   
Blundell, Beasley, \& Bicknell (2003) have inferred 
a Doppler boosting factor of $\sim 10$ in the RQQ PG 1407+263.
This is consistent with the rapid variability found in some
RQQs in this study, and suggests that superluminal motion exists and 
should be measurable in RQQs.  
With the sensitivity now offered by the VLBA in combination with the 
large collecting area of Arecibo, the VLA, and the GBT,
a multi-epoch experiment examining a number of RQQs at high 
resolution and to deep flux density levels may indeed be possible for the first time.  
The discovery of superluminal motion in the cores of RQQs would go a 
long way toward confirming the general thesis suggested here of a 
unified core emission process covering the full spectrum of radio power 
in quasars.   

\acknowledgments 
The authors would like to thank the referee, Ken Kellermann, for a positive
and constructive report.  

{}

\newpage
\begin{deluxetable}{lcccccrrccrc}
\tablenum{1a}
\tabletypesize{\scriptsize}
\tablewidth{0pt}
\tablecaption{Source List: Target Sources\tablenotemark{a,b}}
\tablehead{
\colhead{Source} 
&\colhead{ RA}
&\colhead{Dec}
&\colhead{ $z$ }
&\colhead{ $m_V$}
&\colhead{ $<S>$}
&\colhead{log $R$}
&\colhead{$V_{RMS}$}
&\colhead{$f_{\rm var}$}
&\colhead{$t_{\rm var}$}
&\colhead{$\alpha$}
&\colhead{Class}
\nl
\colhead{}
&\colhead{(B1950)}
&\colhead{(B1950)}
&\colhead{}
&\colhead{}
&\colhead{(mJy)}
&\colhead{}
&\colhead{(\%)}
&\colhead{(\%)}
&\colhead{(yr)}
&\colhead{}
&\colhead{}
}
\startdata
 V0000$-$0229&23:58:08.17&$-$02:46:05.3& 1.070 &17.48 &   0.8  &  0.35 &    7.2 (5.0)&30&0.2&$-0.70$& RQQ\nl
 P0003$+$158 &00:03:25.00&$+$15:53:07.0& 0.450 &15.96 &  229   &  2.20 &   -0.6 (0.6)&...&...&$+1.10$& RLQ \nl
 P0003$+$199 &00:03:45.00&$+$19:55:30.0& 0.025 &13.75 &   2.3  &$-0.69$&   -0.4 (2.3)&...&...&$-0.59$& RQQ \nl
 L0004$+$0036&00:04:36.26&$+$00:36:45.6& 0.317 &17.79 &   0.9  &  0.52 &    6.5 (3.0)&10&0.2&$-0.89$&RIQ\nl
 L0004$+$0224&00:04:53.24&$+$02:24:29.3& 0.300 &17.33 &   1.1  &  0.43 &   20.3 (2.9)&50&0.2&$-0.59$&RQQ\nl
 P0007$+$106 &00:07:56.70&$+$10:41:48.0& 0.090 &15.9  &   146  &  1.98 &  140.9 (0.7)&1000&1&$+0.37$&RIQ\nl
 L0010$+$0131&00:10:21.75&$+$01:31:09.2& 0.433 &18.70 &   2.9  &  1.39 &   -5.6 (2.2)&...&...&$+0.32$&RIQ\nl
 V0031$-$0136&00:29:03.04&$-$01:52:55.5& 2.380 &18.65 &   7.1  &  1.76 &    6.1 (2.5)&20&0.1&$-0.68$&RIQ\nl
 V0042$-$0254&00:39:43.41&$-$03:10:48.3& 2.740 &18.50 &  30.6  &  2.34 &    2.9 (0.8)&10&2&$-0.04$&RLQ\nl
 P0044$+$030 &00:44:31.20&$+$03:03:35.0& 0.624 &15.97 &  22.8  &  1.20 &    9.4 (0.6)&20&2&$-0.14$&RIQ\nl
 P0049$+$171 &00:49:16.50&$+$17:09:41.0& 0.064 &15.88 &   0.7  &$-0.35$&    9.1 (3.6)&20&0.5&$+0.46$&RQQ\nl
 P0052$+$251 &00:52:11.10&$+$25:09:24.0& 0.155 &15.42 &   0.7  &$-0.53$&    9.3 (3.3)&20&0.3&$+0.48$&RQQ\nl
 L0053$+$0124&00:53:00.88&$+$01:24:21.7& 0.440 &17.79 &   4.3  &  1.20 &    6.2 (2.1)&...&...&$-0.13$&RIQ\nl
 V0058$+$1553&00:55:23.46&$+$15:37:03.3& 1.260 &18.40 &  16.6  &  2.04 &    5.9 (1.1)&20&2&$+0.48$&RLQ\nl
 V0152$-$0129&01:50:26.26&$-$01:44:25.5& 2.020 &19.00 &  19.3  &  2.33 &    4.1 (0.9)&10&2&$-0.53$&RLQ\nl
 V0157$+$0138&01:54:34.02&$+$01:23:47.1& 2.130 &17.97 &   6.4  &  1.45 &    9.3 (2.5)&10&2&$+0.11$&RIQ\nl
 P0157$+$001 &01:57:16.30&$+$00:09:10.0& 0.164 &15.20 &   4.6  &  0.19 &    6.2 (2.2)&...&...&$-0.87$&RQQ\nl
 L0256$-$0000&02:56:31.80&$-$00:00:33.6& 3.364 &18.22 &  10.9  &  1.78 &    7.4 (2.0)&20&0.2&$+0.99$&RIQ\nl
 L0302$-$0019&03:02:16.29&$-$00:19:51.8& 3.281 &17.78 &   0.4  &  0.17 &   -7.6 (5.1)&...&...&$+0.19$&RQQ\nl
 V0316$+$0137&03:13:56.23&$+$01:26:29.9& 0.960 &18.40 &   8.6  &  1.75 &    8.4 (2.5)&30&0.2&$+0.69$&RIQ\nl
 P2112$+$059 &21:12:23.60&$+$05:55:12.0& 0.466 &15.52 &   0.5  &$-0.64$&   -6.6 (5.0)&...&...&$-0.52$&RQQ\nl
 P2130$+$099 &21:30:01.30&$+$09:54:59.0& 0.061 &14.62 &   1.5  &$-0.52$&    6.4 (3.2)&...&...&$-0.77$&RQQ\nl
 P2209$+$184 &22:09:30.20&$+$18:27:01.0& 0.070 &15.86 & 167    &  2.02 &   16.2 (0.7)&20&0.1&$+0.66$&RLQ\nl
 L2231$-$0015&22:31:35.13&$-$00:15:29.2& 3.015 &17.53 &   0.5  &  0.16 &   12.5 (4.2)&20&0.5&$+0.65$&RQQ\nl
 L2235$+$0054&22:35:00.82&$+$00:54:58.4& 0.529 &18.55 &   1.0  &  0.87 &   -5.2 (3.1)&...&...&$-0.51$&RIQ\nl
 P2304$+$042 &23:04:30.10&$+$04:16:41.0& 0.042 &15.44 &   1.0  &$-0.37$&   17.9 (2.9)&100&0.5&$+0.67$&RQQ\nl
 P2308$+$098 &23:08:46.50&$+$09:51:55.0& 0.432 &16.12 &  86.1  &  1.83 &    5.4 (0.6)&20&0.2&$-1.94$&RIQ\nl
 P2344$+$092 &23:44:03.70&$+$09:14:05.0& 0.677 &16.00 &1431    &  3.01 &    3.0 (0.6)&10&2&$+0.40$&RLQ\nl
 L2348$+$0210&23:48:23.88&$+$02:10:59.2& 0.504 &18.35 &  19.8  &  2.09 &    7.5 (0.7)&20&2&$-0.49$&RLQ\nl
 L2351$-$0036&23:51:35.36&$-$00:36:29.9& 0.460 &18.47 & 319    &  3.34 &    3.4 (0.6)&10&2&$-0.32$&RLQ\nl

\enddata
\tablenotetext{a} {Source names indicate sample from which
they were chosen.  V = Veron-Cetty and Veron sample, observed previously in
the radio by Bischoff \& Becker (1997); P = Palomar-Green Bright
Quasar Sample, observed by Kellermann et al (1989); L = Large Bright
Quasar Survey, observed by Hooper et al (1996).}

\tablenotetext{b} {The column headings are: $z= $ redshift; $m_{V}$
= optical $V$ magnitude; $<S>$ = mean 8.4 GHz flux density derived from the
measurments reported here; log $R$ = logarithm of the ratio of mean
8.4 GHz radio to optical flux densities; $V_{RMS}$ = fractional debiased RMS
variations at X-band (8.48 GHz), in percent; $f_{var}$ and $t_{var}$ are rough estimations of 
the percentage change and timescale of the strongest variability "event" seen in a given source;
$\alpha = $ radio spectral index between C- and X-bands.  }

\end{deluxetable}

\newpage

\hoffset 0.0cm
\begin{deluxetable}{lcccccrrccrc}
\tablenum{1b}
\tabletypesize{\scriptsize}
\tablewidth{0pt}
\tablecaption{Source List: Calibrators\tablenotemark{a}}
\tablehead{
\colhead{Source}
&\colhead{ RA}
&\colhead{Dec}
&\colhead{ $z$ }
&\colhead{ $m_V$}
&\colhead{ $<S>$}
&\colhead{log $R$}
&\colhead{$V_{RMS}$}
&\colhead{$f_{\rm var}$}
&\colhead{$t_{\rm var}$}
&\colhead{$\alpha$} 
&\colhead{Class}
\nl
\colhead{}
&\colhead{(B1950)}
&\colhead{(B1950)}
&\colhead{}
&\colhead{}
&\colhead{(mJy)}
&\colhead{}
&\colhead{(\%)}
&\colhead{(\%)}
&\colhead{(yr)}
&\colhead{}
&\colhead{}
}
\startdata
0007$+$171 &00:07:59.38&$+$17:07:37.5 &1.601 & 18.0 &   872&3.59&   1.4 (0.7)&...&...&$-0.31$&RLQ \nl
0013$-$005 &00:13:37.35&$-$00:31:52.5 &1.575 & 20.8 &   521&4.49&   6.2 (0.7)&20&2&$-0.32$&RLQ \nl
0019$+$058 &00:19:58.02&$+$05:51:26.5 & ...  & 19.2 &   293&3.59&  13.5 (1.1)&20&0.2& +0.06 &RLQ \nl
0038$-$020 &00:38:24.23&$-$02:02:59.3 &1.178 & 18.5 &   551&3.59&   2.3 (0.8)&10&0.7& +0.07 &RLQ \nl
0039$+$230 &00:39:25.71&$+$23:03:34.8 & ...  & 21.0 &   700&4.70&   0.7 (0.7)&...&...&$-0.43$&RLQ \nl
0109$+$224 &01:09:23.60&$+$22:28:44.1 & ...  & 15.6 &  1138&2.74&  24.4 (0.8)&40&0.4&  0.04 &RLQ \nl
0111$+$021 &01:11:08.57&$+$02:06:24.8 &0.047 & 16.0 &   697&2.70&   2.4 (0.7)&10&1&$-0.07$&RLQ \nl
0122$-$003 &01:22:55.17&$-$00:21:31.2 &1.070 & 17.0 &  1641&3.47&   0.4 (0.8)&...&...& +0.26 &RLQ  \nl
0138$-$097 &01:38:56.85&$-$09:43:51.6 &0.733 & 17.5 &   619&3.24&  12.2 (0.8)&10&30& +0.22 &RLQ \nl
0213$-$026 &02:13:09.87&$-$02:36:51.6 &1.178 & 21.0 &   641&4.66&  -1.4 (0.8)&...&...&  0.03 &RLQ  \nl
0216$+$011 &02:16:32.45&$+$01:07:13.3 &1.623 & 20.0 &   629&4.25&  -1.7 (1.0)&...&...&  0.03 &RLQ \nl
0237$-$027 &02:37:13.71&$-$02:47:32.8 &1.116 & 21.0 &   445&4.50&  26.0 (0.7)&10&0.5& +0.59 &RLQ\nl
0256$+$075 &02:56:46.99&$+$07:35:45.2 &0.893 & 18.0 &   766&3.54&  12.2 (0.9)&40&2&$-0.12$&RLQ\nl
2121$+$053 &21:21:14.80&$+$05:22:27.4 &1.878 & 17.5 &  1219&3.54&  35.7 (0.8)&50&0.5& +0.37 &RLQ\nl
2150$+$173 &21:50:02.23&$+$17:20:29.8 & ...  & 21.0 &   751&4.73&   8.4 (0.8)&20&0.3& +0.14 &RLQ\nl
2239$+$096 &22:39:19.84&$+$09:38:09.9 &1.707 & 19.5 &   532&3.98&   3.5 (1.0)&10&0.5& +0.10 &RLQ\nl
2254$+$024 &22:54:44.61&$+$02:27:14.2 &2.081 & 18.0 &   399&3.25&   2.2 (0.7)&...&...& +0.84 &RLQ\nl
2318$+$049 &23:18:12.13&$+$04:57:23.4 &0.662 & 19.0 &  1005&4.05&   9.4 (0.7)&10&0.5&$-0.06$&RLQ\nl
2328$+$107 &23:28:08.78&$+$10:43:45.5 &1.489 & 18.1 &  1052&3.71&  -2.1 (1.0)&...&...&$-0.18$&RLQ\nl
2344$+$092 &23:44:03.77&$+$09:14:05.4 &0.667 & 15.6 &  1440&2.85&  -1.1 (0.9)&...&...&$-0.13$&RLQ\nl

\enddata

\tablenotetext{a} {Column headings as in Table 1a.}

\end{deluxetable}


\clearpage
\newpage

\hoffset 0.0cm
\begin{deluxetable}{lccc}
\tablenum{2}
\tablewidth{3.7in}
\tablecaption{Log of Observations}
\tablehead{
\colhead{Date} 
&\colhead{Band} 
&\colhead{VLA Configuration}
&\colhead{Beam size\tablenotemark{1}}
}
\startdata
1/17/97&X&A/B Hybrid&$1.3'' \times 1.0''$\nl
3/11/97&X&B&$1.3'' \times 1.0''$\nl
4/03/97\tablenotemark{2}&X&B&$1.3'' \times 1.0''$\nl
5/04/97&X&B&$1.3'' \times 1.0''$\nl
5/30/97&X&B/C Hybrid&$3.0'' \times1.3''$\nl
6/29/97&X&C&$3.0'' \times 3.0''$\nl
8/07/97&X&C/D Hybrid&$3.5'' \times 3.0''$\nl
8/21/97&X&C/D Hybrid&$3.5'' \times 3.0''$\nl
9/20/97&X&C/D Hybrid&$3.5'' \times 3.0''$\nl
4/15/99&X&D&$9'' \times 8''$\nl
6/03/99&C&A/D Hybrid&$15'' \times 10''$\nl
6/03/99&L&A/D Hybrid&$45''\times 30''$\nl
\enddata
\tablenotetext{1} {Beam sizes differ slightly for each source; listed
values approximate. All UV data for the first epoch tapered by 160 k$\lambda$.} 
\tablenotetext{2} {Phases noisy; only strong sources reported.} 
\end{deluxetable}

\newpage

\begin{deluxetable}{lrrrrrrrrrrrr}
\scriptsize
\tablenum{3a}
\tabletypesize{\scriptsize}
\setlength{\tabcolsep}{0.03in}
\tablewidth{0pt}
\tablecaption{Target Source Flux Densities (mJy/Beam)\tablenotemark{a} }
\tablehead{
\colhead{Source} &\colhead{1/17/97\tablenotemark{b}}
&\colhead{3/11/97}  
&\colhead{4/3/97}
&\colhead{5/4/97}
&\colhead{5/30/97}
&\colhead{6/29/97}
&\colhead{8/7/97}
&\colhead{8/21/97}
&\colhead{9/20/97}
&\colhead{4/15/99}
}
\startdata

V0000$-$0229&     ...      &  ...         &  ...         &  ...         &   0.85 (0.09)&   0.89 (0.09)&   0.62 (0.09)&   0.66 (0.10)&
0.88 (0.11)&   0.71 (0.08) \nl
P0003+158   &    226 (5)   & 226 (5)      & 231 (5)      & 231 (5)      & 230 (5)      & 229 (5)      & 230 (5)      & 229.4 (5)& 221 (5)      & 239 (5)        \nl
P0003+199   &   2.10 (0.15)&   2.40 (0.16)&    ...   &   2.11 (0.14)&   2.36 (0.16)&   2.51 (0.17)&   2.40 (0.16)&   2.35 (0.16)&
2.03 (0.15)&   2.25 (0.15) \nl
L0004+0036  &   1.08 (0.11)&   0.86 (0.08)&    ...   &   0.96 (0.08)&   1.07 (0.08)&   1.02 (0.08)&   0.90 (0.09)&   0.83 (0.08)&
0.82 (0.08)&   0.80 (0.07) \nl
L0004+0224  &   0.84 (0.12)&   0.79 (0.07)&    ...   &   0.69 (0.06)&   1.02 (0.09)&   1.34 (0.10)&   1.13 (0.09)&   1.29 (0.11)&
1.23 (0.10)&   1.36 (0.10) \nl
P0007+106   &  44.2 (1.7)&  42.7 (1.2)&  45.3 (1.4)&  52.1 (1.2)&  61.7 (1.4)&  79.5 (1.7)& 108 (2)& 123 (3)&
147 (3)& 752 (15) \nl
L0010+0131  &   2.94 (0.21)&   2.91 (0.19)&    ...   &   2.70 (0.18)&   2.80 (0.18)&   3.03 (0.19)&   2.75 (0.18)&   2.84 (0.18)&
2.82 (0.18)&   2.92 (0.18) \nl
V0031$-$0136&     ...      &     ...      &     ...      &     ...      &   7.74 (0.47)&   7.59 (0.46)&   6.98 (0.43)&   7.52 (0.46)&
6.09 (0.38)&   6.47 (0.39) \nl
V0042$-$0254&     ...      &     ...      &     ...      &     ...      &  30.2 (0.6)&  31.7 (0.6)&  31.2 (0.6)&  30.7 (0.6)&
31.2 (0.6)&  28.4 (0.6) \nl
P0044+030   &  21.0 (0.5)&  20.3 (0.4)&  23.8 (0.5)&  21.8 (0.5)&  21.3 (0.4)&  23.3 (0.5)&  23.7 (0.5)&  21.8 (0.5)&
22.4 (0.5)&  28.5 (0.6) \nl
P0049+171   &   0.56 (0.10)&   0.69 (0.06)&    ...   &   0.75 (0.07)&   0.71 (0.07)&   0.76 (0.07)&   0.63 (0.06)&   0.83 (0.09)&
0.53 (0.08)&   0.81 (0.08) \nl
P0052+251   &   0.58 (0.10)&   0.67 (0.06)&    ...   &   0.74 (0.07)&   0.82 (0.06)&   0.77 (0.07)&   0.68 (0.07)&   0.71 (0.07)&
0.73 (0.07)&   0.96 (0.08) \nl
L0053+0124  &   3.77 (0.24)&   4.02 (0.25)&    ...   &   3.79 (0.24)&   4.18 (0.26)&   4.50 (0.28)&   4.65 (0.29)&   5.01 (0.32)&
4.39 (0.29)&   4.32 (0.27) \nl
V0058+1553  &     ...      &     ...      &     ...      &     ...      &  17.2 (0.4)&  16.9 (0.4)&  16.8 (0.4)&  17.6 (0.4)&
17.1 (0.4)&  14.2 (0.9) \nl
V0152$-$0129&     ...      &     ...      &     ...      &     ...      &  19.1 (0.4)&  19.8 (0.4)&  19.7 (0.4)&  19.8 (0.4)&
19.8 (0.4)&  17.4 (0.4) \nl
V0157+0138  &     ...      &     ...      &     ...      &     ...      &   6.45 (0.39)&   6.76 (0.41)&   5.82 (0.39)&   5.96 (0.36)&
5.53 (0.34)&   7.67 (0.47) \nl
P0157+001   &   4.11 (0.26)&   3.92 (0.25)&    ...   &   4.68 (0.39)&   4.78 (0.30)&   4.96 (0.31)&   4.22 (0.29)&   4.60 (0.29)&
4.79 (0.30)&   5.33 (0.34) \nl
L0256$-$0000&   9.97 (0.61)&  10.75 (0.65)&    ...   &  10.09 (0.61)&   9.71 (0.59)&  11.65 (0.70)&  10.97 (0.66)&  11.06 (0.67)&
10.81 (0.66)&  13.43 (0.81) \nl
L0302$-$0019&   0.40 (0.08)&   0.33 (0.05)&    ...   &   0.44 (0.06)&   0.36 (0.05)&   0.44 (0.06)&   0.45 (0.06)&   0.37 (0.06)&
0.45 (0.07)&   0.30 (0.05) \nl
V0316+0137  &     ...      &     ...      &     ...      &     ...      &   9.09 (0.55)&   9.60 (0.58)&   9.32 (0.57)&   8.57 (0.52)&
6.94 (0.44)&   8.08 (0.49) \nl
P2112+059   &     ...      &     ...      &     ...      &     ...      &   0.55 (0.06)&   0.60 (0.07)&   0.44 (0.07)&   0.48 (0.06)&
0.49 (0.06)&   0.55 (0.06) \nl
P2130+099   &     ...      &     ...      &     ...      &     ...      &     ...      &   1.36 (0.10)&   1.43 (0.10)&   1.33 (0.10)&
1.47 (0.11)&   1.73 (0.12) \nl
P2209+184   & 203 (4)& 202 (4)& 174 (4)&    ...   & 155 (3)& 145 (3)& 149 (3)& 159 (3)&
200    (4)& 121 (3) \nl
L2231$-$0015&   0.43 (0.08)&   0.43 (0.05)&    ...   &   0.59 (0.06)&   0.55 (0.06)&   0.60 (0.07)&   0.68 (0.07)&   0.48 (0.07)&
0.38 (0.06)&   0.50 (0.06) \nl
L2235+0054  &   0.84 (0.11)&    ...   &    ...   &   1.01 (0.08)&   1.03 (0.09)&   1.04 (0.09)&   0.91 (0.07)&   1.07 (0.09)&
0.96 (0.08)&   0.99 (0.08) \nl
P2304+042   &   0.65 (0.10)&   0.98 (0.08)&    ...   &   1.16 (0.08)&   1.25 (0.09)&   1.15 (0.09)&   0.94 (0.08)&   0.88 (0.08)&
0.73 (0.07)&   1.15 (0.09) \nl
P2308+098   &  83.3 (1.7)&  85.4 (1.7)&  86.3 (1.7)&  84.1 (1.7)&  86.4 (1.7)&  83.4 (1.7)&  81.2 (1.6)&  84.4 (1.7)&
86.0 (1.7)& 100.3 (2.1) \nl
P2344+092   &1434 (29)&1428 (29)&1461 (29)&1438 (29)&1428 (29)&1450 (29)&1467 (29)&1462
(29)&1460 (29)&1283 (26) \nl
L2348+0210  &  22.3 (0.5)  &  20.4 (0.4)  &  19.6 (0.4)  &  21.1 (0.4)  &  19.0 (0.4)  &  19.5 (0.4)  &  19.9 (0.4)  &  19.8 (0.4)  &
20.1 (0.4)  &  16.0 (0.3)    \nl
L2351$-$0036& 333 (7)     & 325 (7)     & 334 (7)     & 324 (7)     & 321 (7)     & 318 (6)     & 320 (6)     & 318 (6)     &
319 (6)     & 286 (6)        \nl
\enddata

\tablenotetext{a} {NOTE: All source flux densities derived from peak map values.  Errors in parentheses
represent statistical and calibration uncertainties added in quadrature.
The calibration uncertainties are 2\% for flux densities greater than 15 mJy, and 6\% for flux densities less than 15 mJy
(see text).}

\tablenotetext{b} {Observations for 1/17/97 done in hybrid AB configuration.  Maps were made using a taper of
160 k$\lambda$, resulting in a synthesized beamsize of $\approx 1.3 \times 1.0\arcsec$, in order
to match as closely as possible the  second epoch measurements (3/11/97) and reduce resolution
effects.  See Table 1 for configuration details for other epochs. }

\end{deluxetable}

\newpage

\hoffset -0.0cm
\tableheadfrac{0.05}
\begin{deluxetable}{lrrrrrrrrrrrr}
\scriptsize
\tablenum{3b}
\tabletypesize{\scriptsize}
\setlength{\tabcolsep}{0.06in}
\tablewidth{0pt}
\tablecaption{Calibrator Flux Densities (mJy/Beam)\tablenotemark{a}}
\tablehead{
\colhead{Source} &\colhead{1/17/97}
&\colhead{3/11/97}  
&\colhead{4/3/97}
&\colhead{5/4/97}
&\colhead{5/30/97}
&\colhead{6/29/97}
&\colhead{8/7/97}
&\colhead{8/21/97}
&\colhead{9/20/97}
&\colhead{4/15/99}
}
\startdata
0007+171 & 895 (19) & 898 (21) & 902 (21) & 882 (19) & 874 (18) & 863 (20) & 867 (17) & 864 (17) & 829 (17) & 845 (17) \nl
0013$-$005 &    ...  & 515 (12) & 503 (12) & 501 (11) & 495 (10) & 501 (11) & 506 (10) & 520 (10) & 534 (11) & 613 (12) \nl
0019+058 & 258 ( 5) & 360 ( 9) &    ...  & 286 ( 6) & 267 ( 6) &    ...  &    ...  &    ...  &    ...  &    ...  \nl
0038$-$020 & 581 (12) &    ...  &    ...  &    ...  &    ...  & 566 (12) & 547 (11) & 539 (11) & 532 (11) & 540 (11) \nl
0039+230 & 706 (14) & 689 (16) & 693 (16) & 692 (14) & 687 (14) & 709 (15) & 721 (14) & 717 (14) & 715 (14) & 671 (13) \nl
0109+224 &    ...  &    ...  &    ...  &    ...  &1053 (22) &1239 (25) &1274 (26) &1314 (26) &1389 (28) & 558 (11) \nl
0111+021 &703 (14) & 715 (17) & 728 (17) & 712 (15) & 712 (15) & 708 (15) & 686 (14) & 684 (14) & 671 (14) & 650 (13) \nl
0122$-$003 &    ...  &    ...  &    ...  &    ...  &1621 (33) &1626 (36) &1635 (33) &1620 (33) &1626 (33) &1717 (35) \nl
0138$-$097 &    ...  &    ...  &    ...  &    ...  & 706 (15) & 690 (15) & 626 (13) & 611 (12) & 615 (12) & 469 ( 9) \nl
0213$-$026 &    ...  &    ...  &    ...  &    ...  & 633 (13) & 640 (13) & 634 (13) & 638 (13) & 641 (13) & 661 (13) \nl
0216+011 & 629 (13) & 615 (15) & 643 (15) & 628 (13) & 631 (13) &    ...  &    ...  &    ...  &    ...  &    ...  \nl
0237$-$027 & 466 (10) & 438 (11) & 442 (10) & 413 ( 9) & 393 ( 8) & 393 ( 8) & 383 ( 8) & 378 ( 8) & 365 ( 7) & 782 (16) \nl
0256+075 &    ...  &    ...  &    ...  &    ...  &    ...  & 839 (17) & 802 (16) & 816 (16) & 794 (16) & 580 (12) \nl
2121+053 &    ...  &    ...  &    ...  &    ...  & 861 (18) & 917 (23) &1056 (21) &1069 (21) &1260 (25) &2151 (43) \nl
2150+173 & 722 (15) & 699 (16) & 641 (16) &    ...  &    ...  & 771 (18) & 763 (15) & 761 (15) & 765 (15) & 884 (18) \nl
2239+096 & 504 (10) & 517 (12) & 522 (13) & 560 (12) & 556 (12) &    ...  &    ...  &    ...  &    ...  &    ...  \nl
2254+024 &    ...  & 411 (10) & 395 ( 9) & 401 ( 9) & 394 ( 8) & 393 ( 9) & 402 ( 8) & 385 ( 8) & 385 ( 8) & 426 ( 9) \nl
2318+049 &1042 (21) &1040 (25) &1062 (24) &1055 (22) &1017 (21) & 976 (21) & 907 (18) & 887 (18) & 864 (17) &1207 (24) \nl
2328+107 &1064 (21) &1044 (25) &1050 (23) &1055 (22) &1049 (22) &    ...  &    ...  &    ...  &    ...  &    ...  \nl
2344+092 &    ...  &    ...  &    ...  &    ...  &    ...  &1445 (32) &1464 (29) &1461 (29) &1435 (29) &1394 (28) \nl
\enddata

\tablenotetext{a} {NOTE: All source flux densities derived from point source model fits to the UV data. 
Errors in parentheses include statistical and calibration uncertainties added in quadrature. 
Calibration uncertainties are assumed to be $\pm 2$\%}


\end{deluxetable}

\hoffset 0.0cm
\begin{deluxetable}{lccccc}
\tablenum{4}
\tablewidth{4.5in}
\tablecaption{Variability and Spectral Index Statistics}
\tablehead{
\colhead{Class}
&\colhead{Number} 
&\colhead{$V^*_{RMS}$\tablenotemark{a}} 
&\colhead{$V^*_{RMS}$} 
&\colhead{$\alpha$}
&\colhead{$\alpha$}
\\
\colhead{}
&\colhead{in Class}
&\colhead{Mean}
&\colhead{Median}
&\colhead{Mean}
&\colhead{Median}
}
\startdata
RQQ & 11 & 8.1 (6.8) & 7.2 & $-0.15~(0.63)$ & $-0.52$ \nl
RIQ & 10/11\tablenotemark{b} &5.9 (3.3) & 6.4 & $-0.17~(0.82)$ & $-0.13$ \nl
RLQ & 8 & 5.4 (4.9) & 3.8 & $+0.16~(0.59)$ & $+0.19$ \nl
CAL & 20 &8.1 (10.1) & 3.0 & $+0.06~(0.30)$ & $+0.04$ \nl
\enddata
\tablenotetext{a} {For statistical comparison the quantity $V^*_{RMS}$ is 
used.  This 
is just the debiased RMS with all negative values set to zero, since
a negative number indicates that a source did not vary
in this data set above the noise. Numbers in 
parentheses are the standard deviations of the values for that class.}
\tablenotetext{b} {$V^*_{RMS}$ for P0007+10 (= III Zw 2) not used in the
calculation of the mean $V^*_{RMS}$ for RIQs because of its extreme value. 
However, the spectral index was used in the corresponding calculation.}
\end{deluxetable}

\newpage

\begin{figure*}
\figurenum{1}
\epsscale{1.2}
\includegraphics[height=18cm, angle=-90]{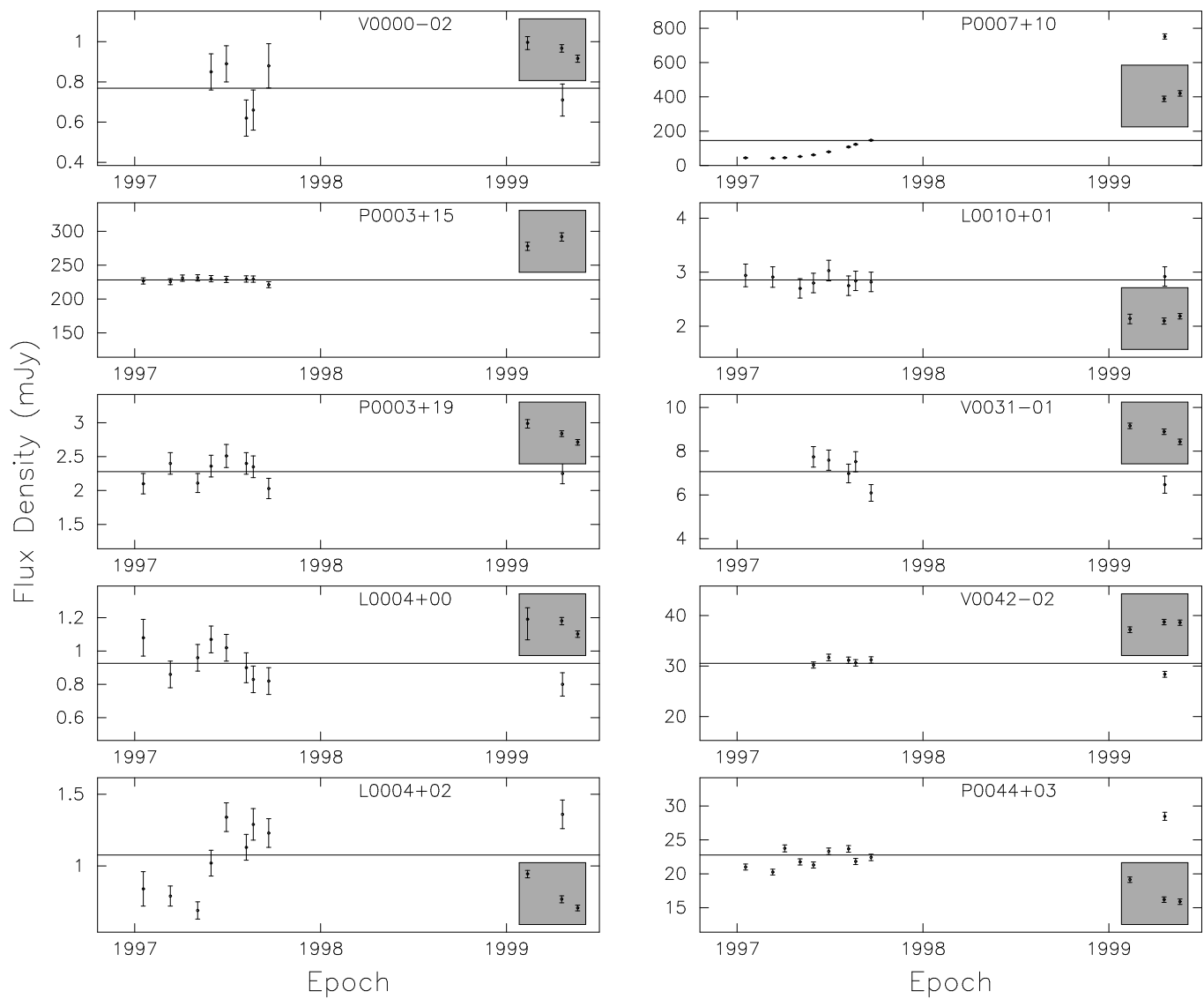} 
\caption{ VLA X-band (8.48 GHz) radio light curves for quasars from the 
target sample.  The 
inset boxes show the radio spectrum at L-, C-, and X-bands (1.49, 4.89, and 
8.48 GHz respectively), going from left to right.  The horizontal
line running through 
 the center of each plot represents the mean value of the X-band flux density
for that source. In most cases the vertical axis is scaled to run from 
0.5 to 1.5 times the mean flux density.
} \end{figure*}

\begin{figure*}
\figurenum{1 (cont)}
\epsscale{1.2}
\includegraphics[height=18cm, angle=-90]{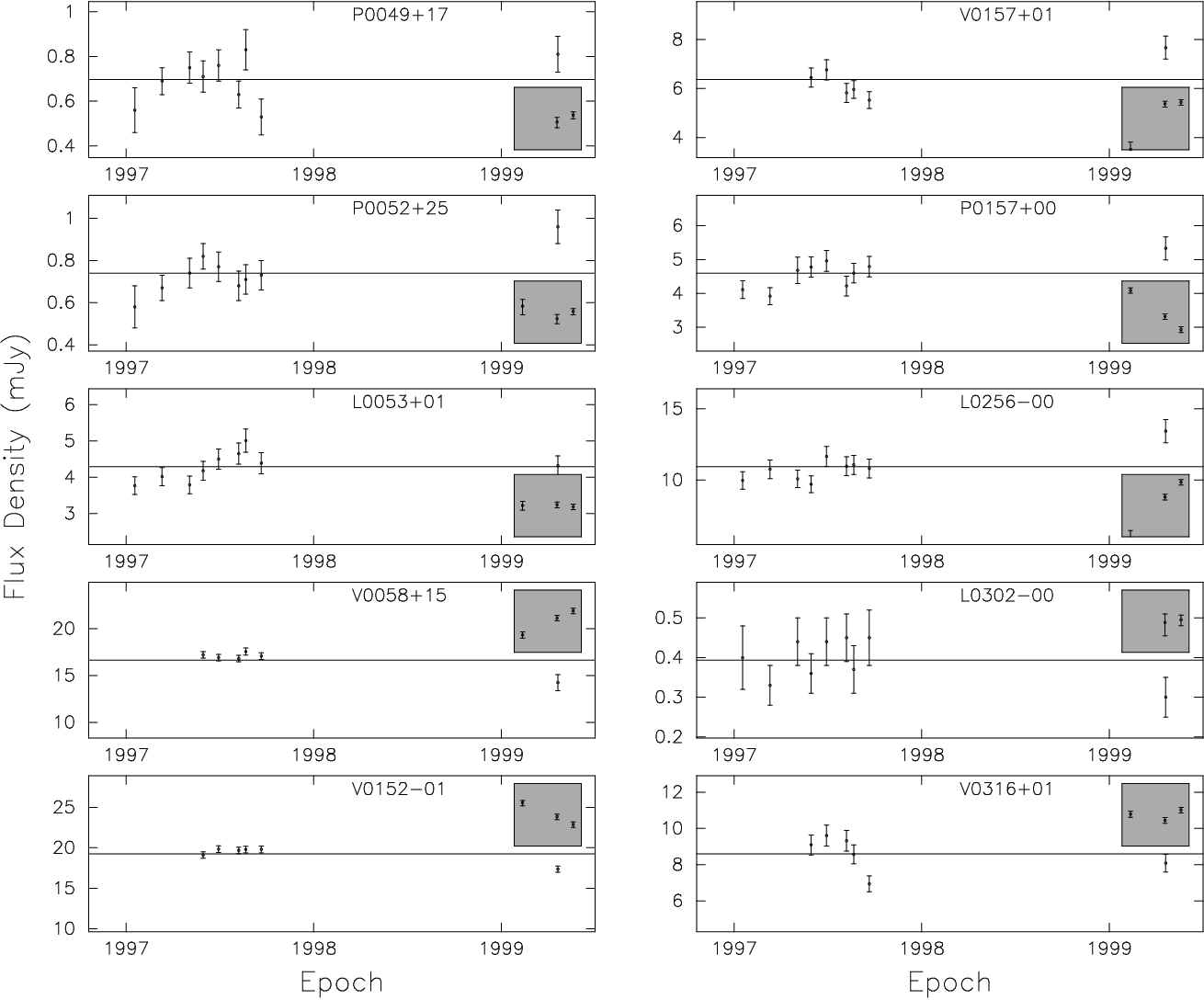} 
\caption{
} \end{figure*}

\begin{figure*}
\figurenum{1(cont)}
\epsscale{1.2}
\includegraphics[height=18cm, angle=-90]{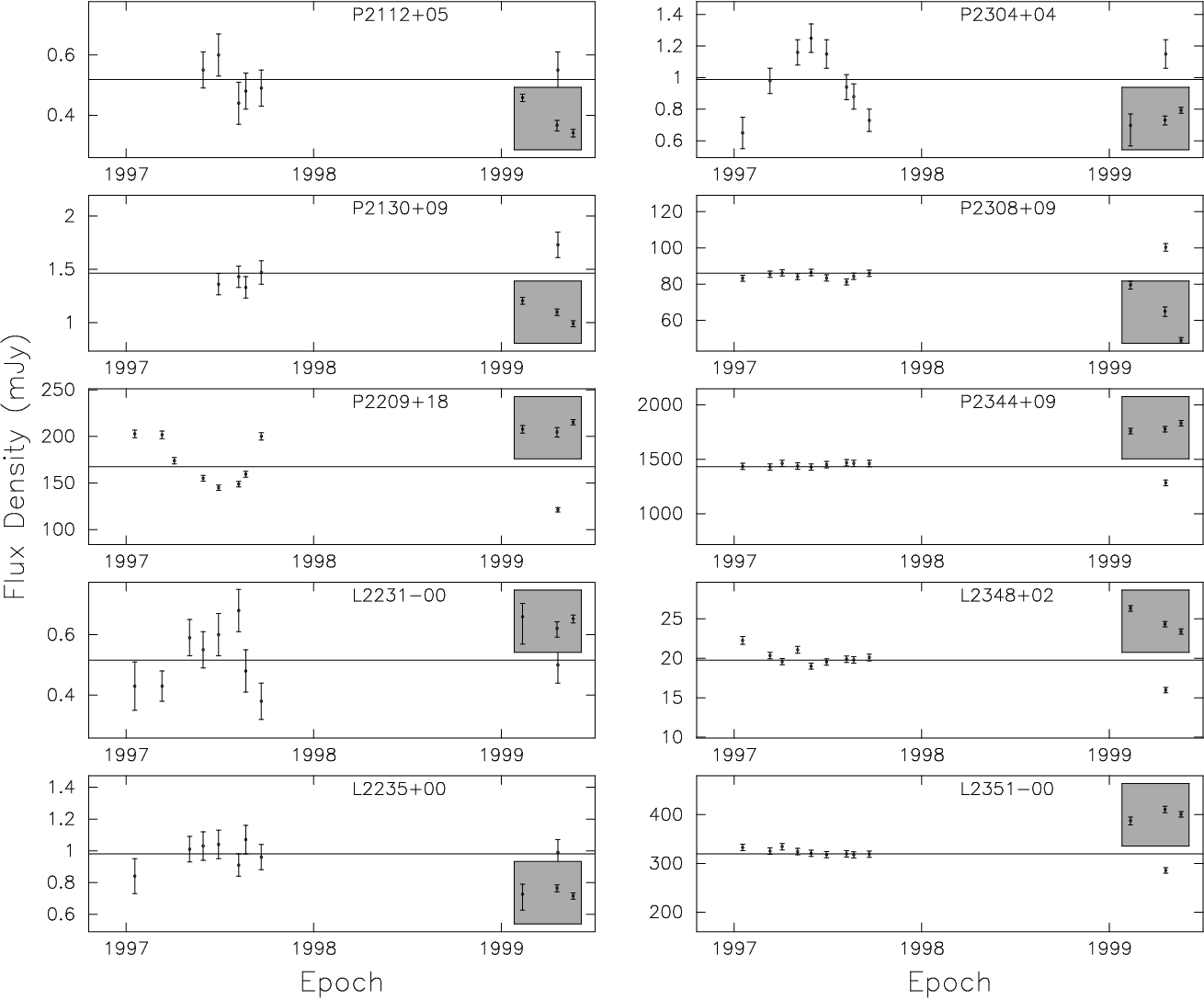} 
\caption{}
\end{figure*}

\begin{figure*}
\figurenum{2}
\epsscale{1.2}
\includegraphics[height=18cm, angle=-90]{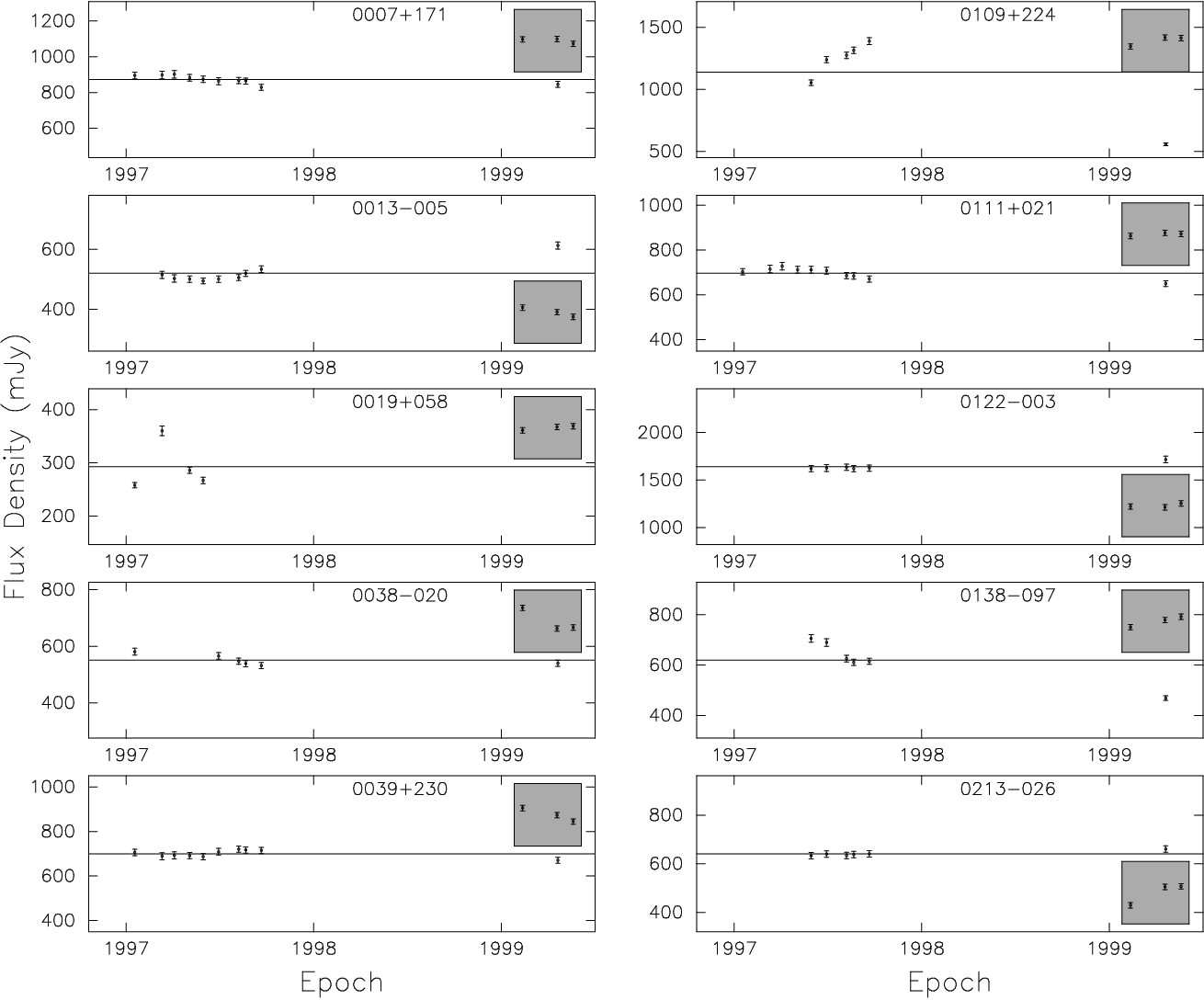} 
\caption{ Same as Figure 1, but for the calibrator sources.}
\end{figure*}

\begin{figure*}
\figurenum{2(cont)}
\epsscale{1.2}
\includegraphics[height=18cm, angle=-90]{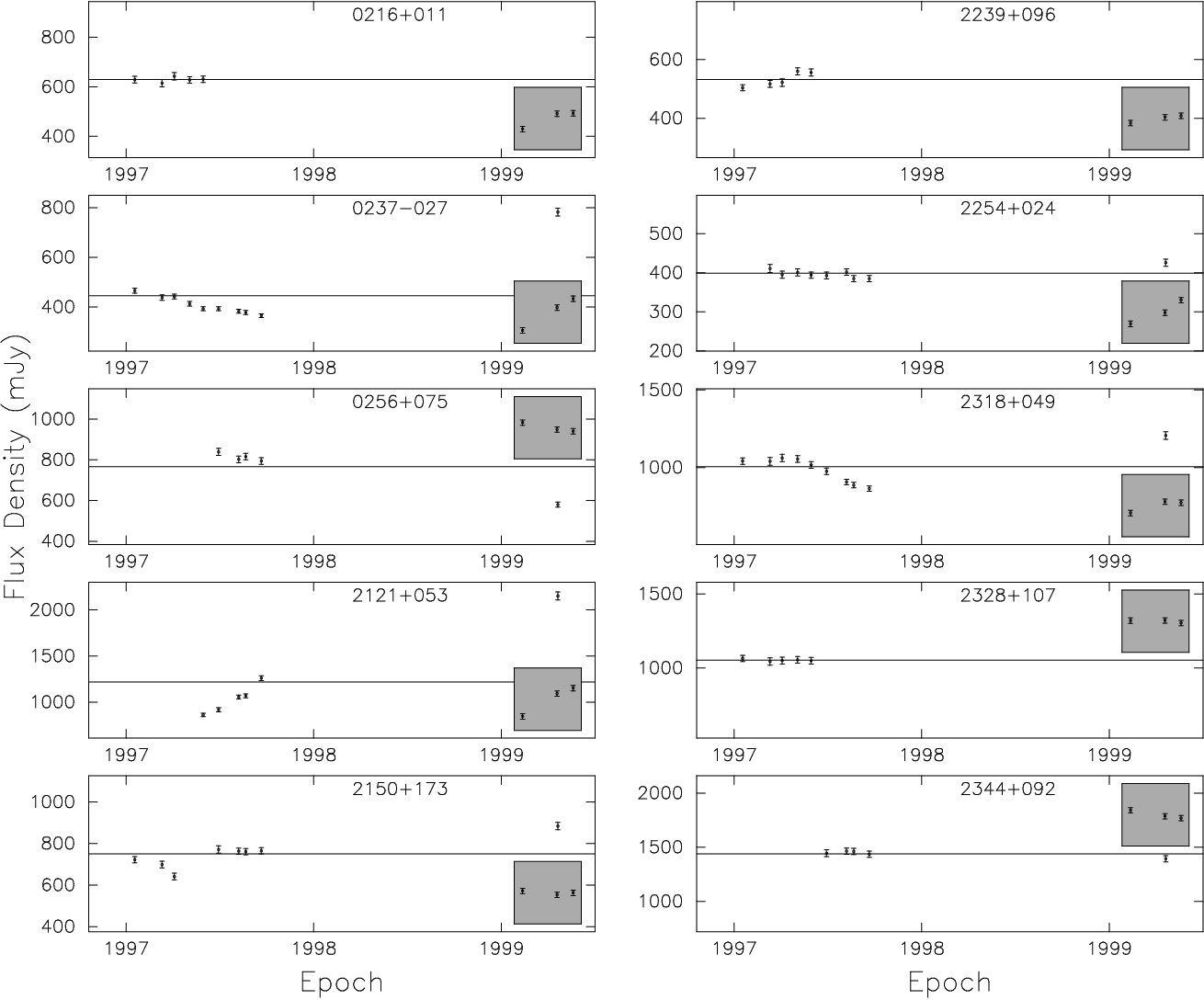} 
\caption{}
\end{figure*}

\begin{figure*}
\figurenum{3}
\epsscale{1.2}
\includegraphics[height=16cm, angle=-90]{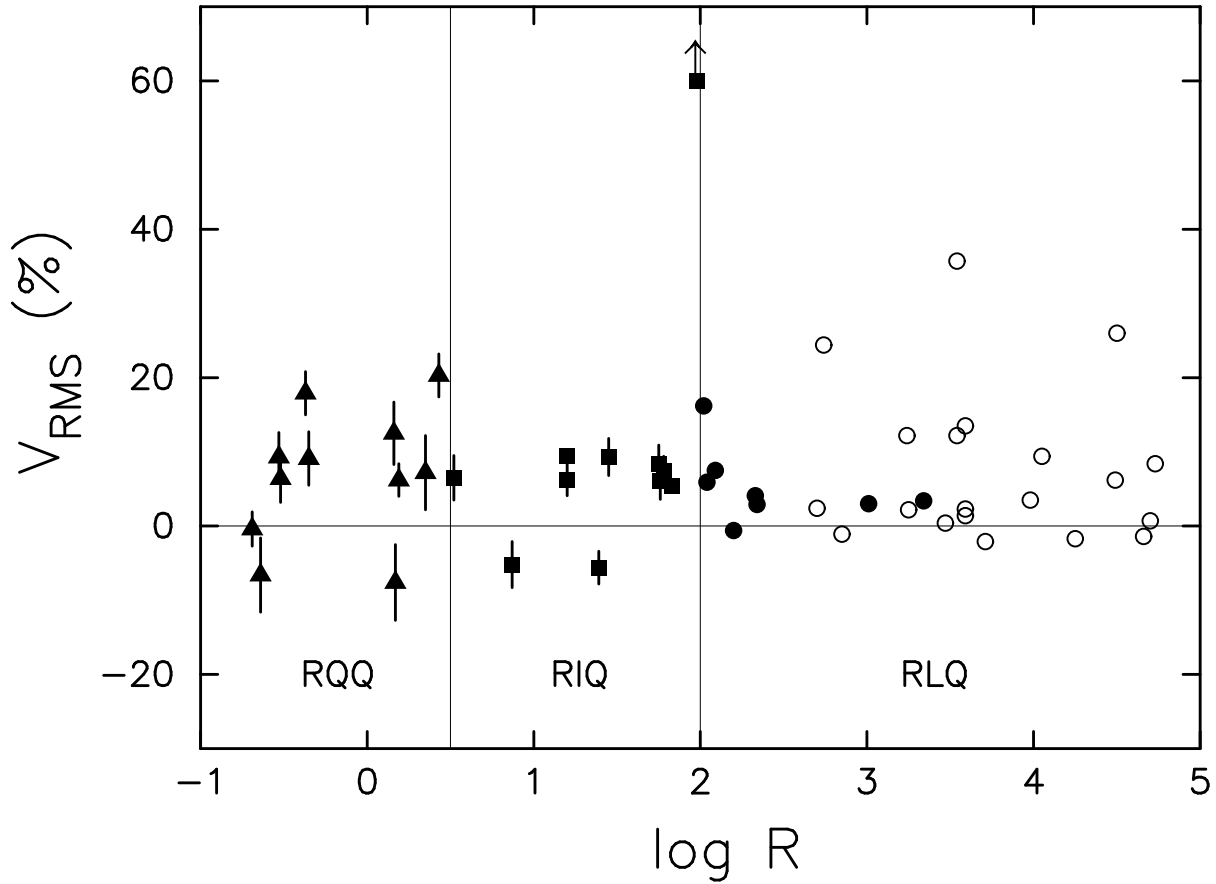} 
\caption{Debiassed variability index $V_{RMS}$, expressed in percent, versus
radio-to-optical flux density ratio $R$.  The plot is divided into regions of radio
quiet, radio intermediate, and radio loud objects. The mean radio flux density over
all epochs is used in the calculation of $R$.  The closed symbols represent
target quasars, while the open circles represent calibrators. The variability 
index of III Zw 2 -- the highest point in the diagram -- is shown here
only as a lower limit (actual value 140.9) to keep the scale of the plot 
in reasonable bounds. We define the variability index to be negative in cases where
the actual fluctuations are smaller than expected from the error bars, due
either to chance or to overestimation of the errors (see
text).}
\end{figure*}

\begin{figure*}
\figurenum{4}
\epsscale{1.2}
\includegraphics[height=16cm, angle=-90]{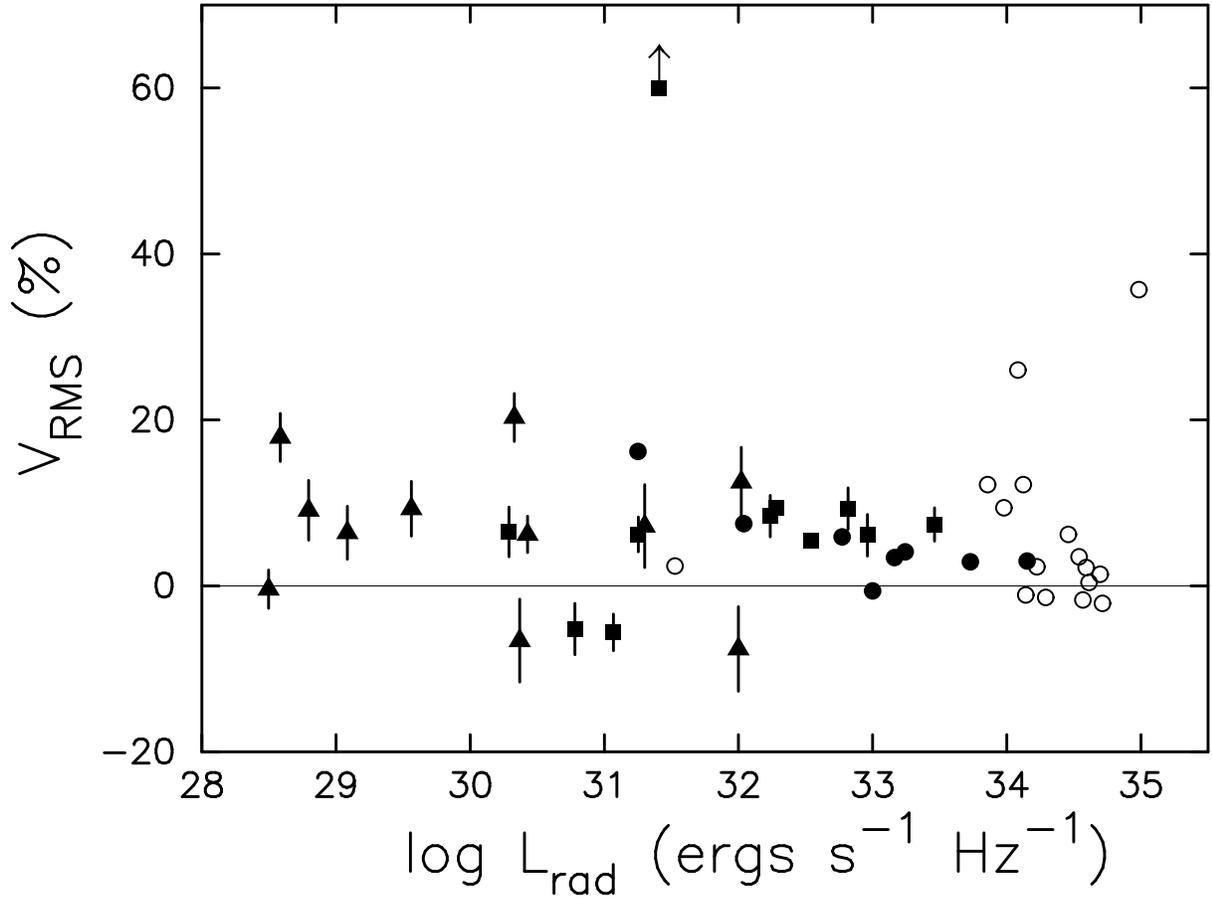} 
\caption{Debiassed variability index $V_{RMS}$ versus radio luminosity.
Symbols are the same as in Figure 3: 
closed triangles for RQQs, 
closed squares for RIQs, closed circles for RLQs, and open circles for 
calibrators.  Four calibrators are not plotted because they lack redshifts.
}
\end{figure*}

\begin{figure*}
\figurenum{5}
\epsscale{1.2}
\includegraphics[height=16cm, angle=-90]{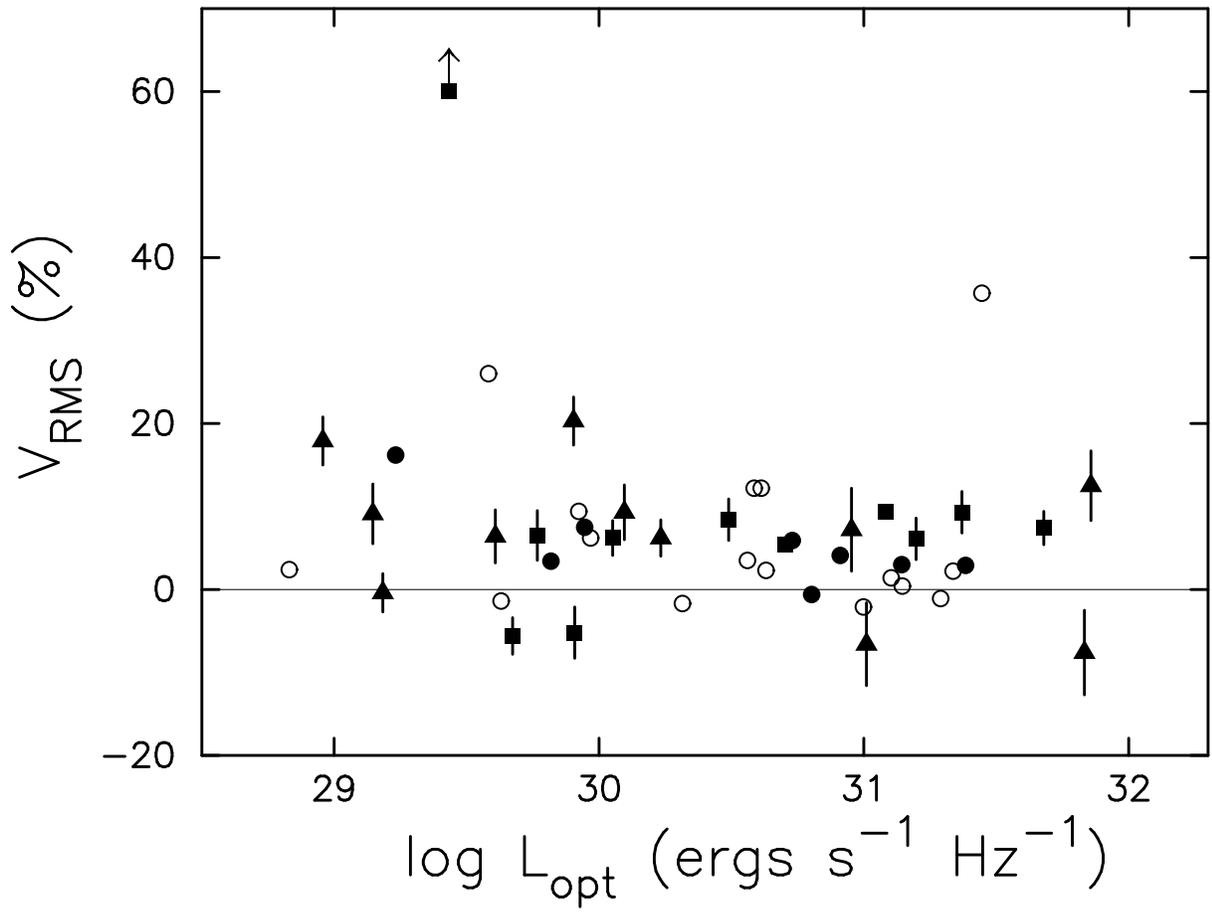} 
\caption{Debiassed variability index $V_{RMS}$ versus optical luminosity.
Symbols same as in Figure 4.}
\end{figure*}

\begin{figure*}
\figurenum{6}
\epsscale{1.2}
\includegraphics[height=16cm, angle=-90]{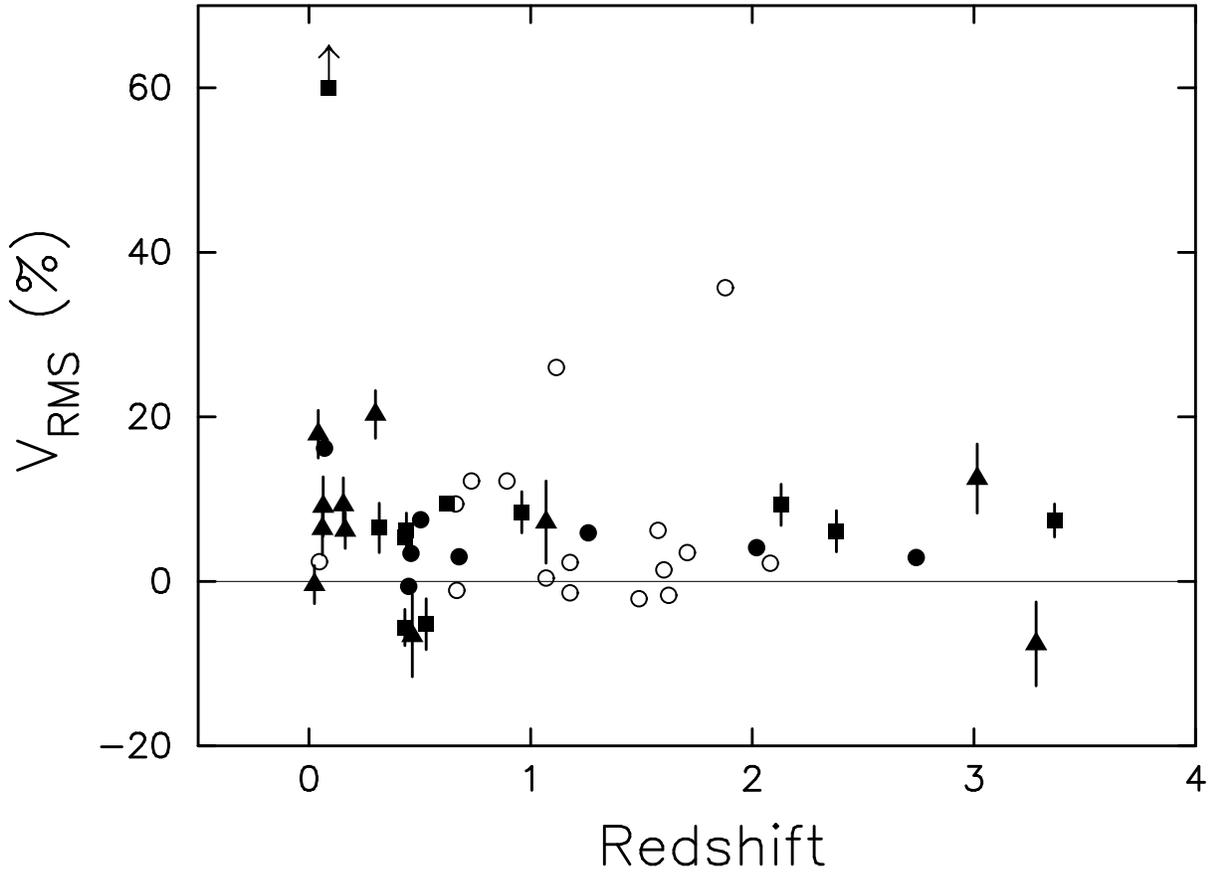} 
\caption{Debiassed variability index $V_{RMS}$ versus redshift.
Symbols same as in Figure 4.}
\end{figure*}

\begin{figure*}
\figurenum{7}
\epsscale{1.2}
\includegraphics[height=16cm, angle=-90]{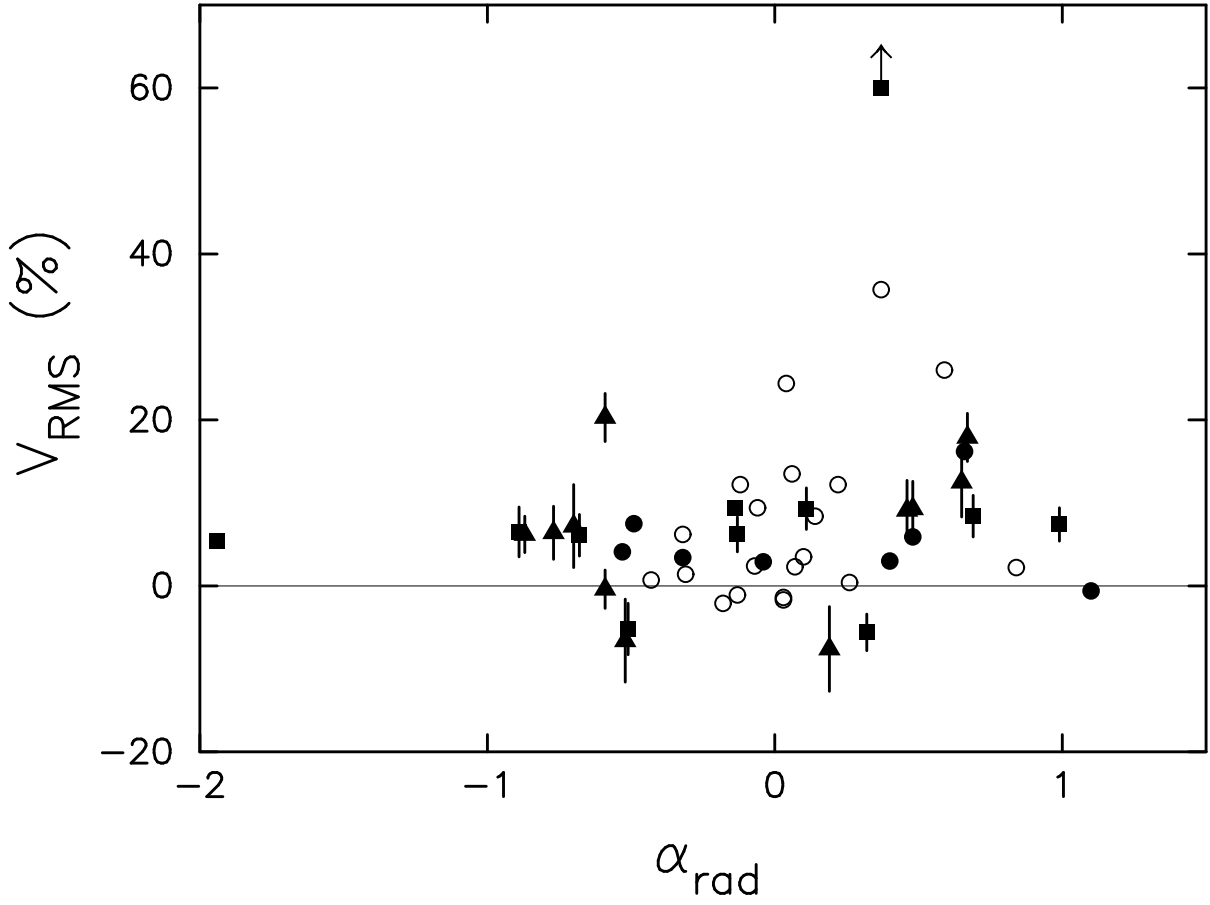} 
\caption{Debiassed variability index $V_{RMS}$ versus radio spectral index
between C- and X-band.
Symbols same as in Figure 4.}
\end{figure*}

\begin{figure*}
\figurenum{8}
\epsscale{1.2}
\includegraphics[height=14cm, angle=-90]{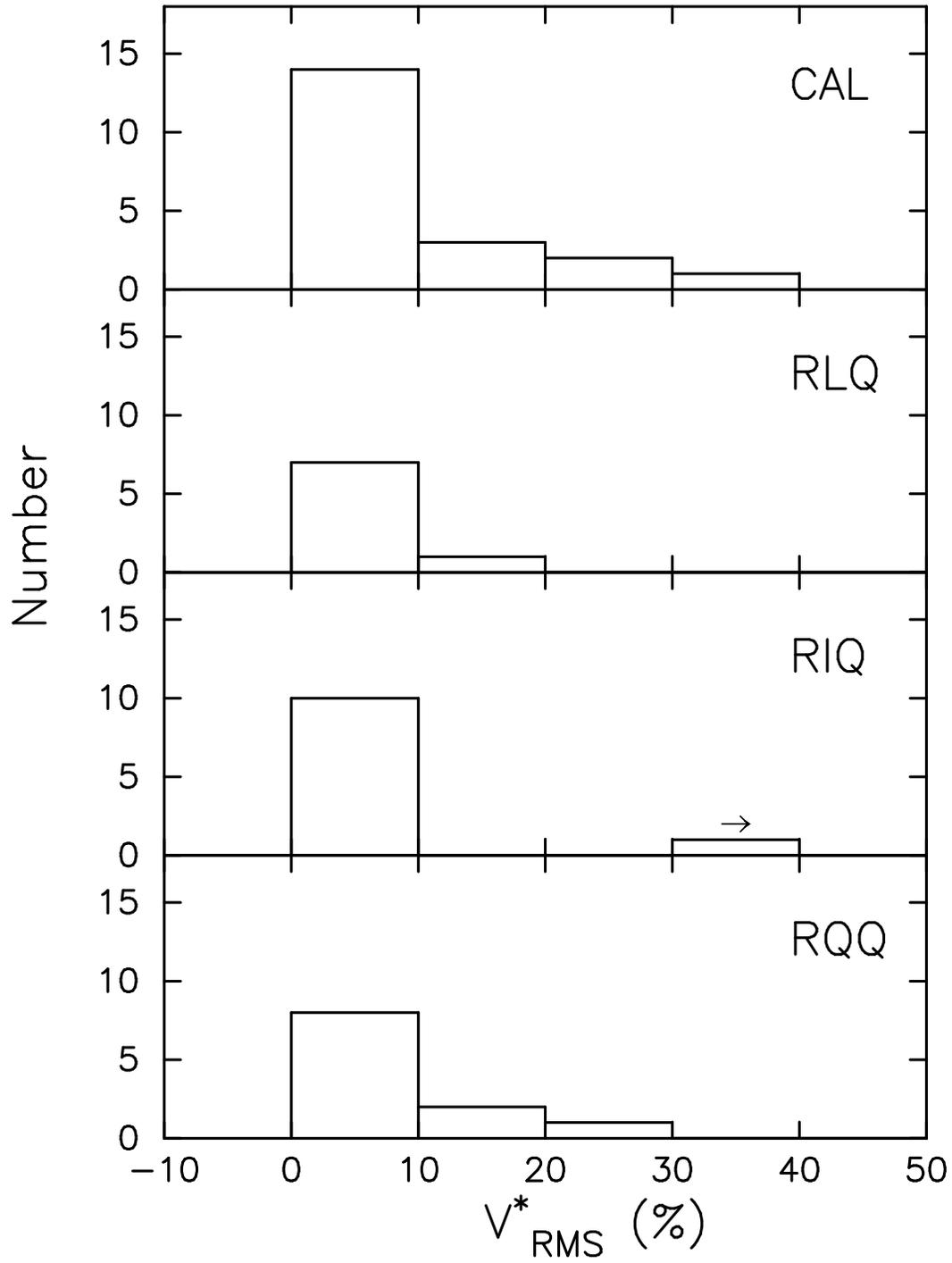} 
\caption{Comparison of histograms of debiased RMS for the four source 
classes. Here we have set negative values to zero, since such values mean
that the sources were not variable.  This is indicated by the asterix in 
$V^{*}_{RMS}$.
Value for III Zw 2 is off the scale of the abscissa (indicated by arrow).
}
\end{figure*}

\clearpage
\begin{figure*}
\figurenum{9}
\epsscale{1.2}
\includegraphics[height=14cm, angle=-90]{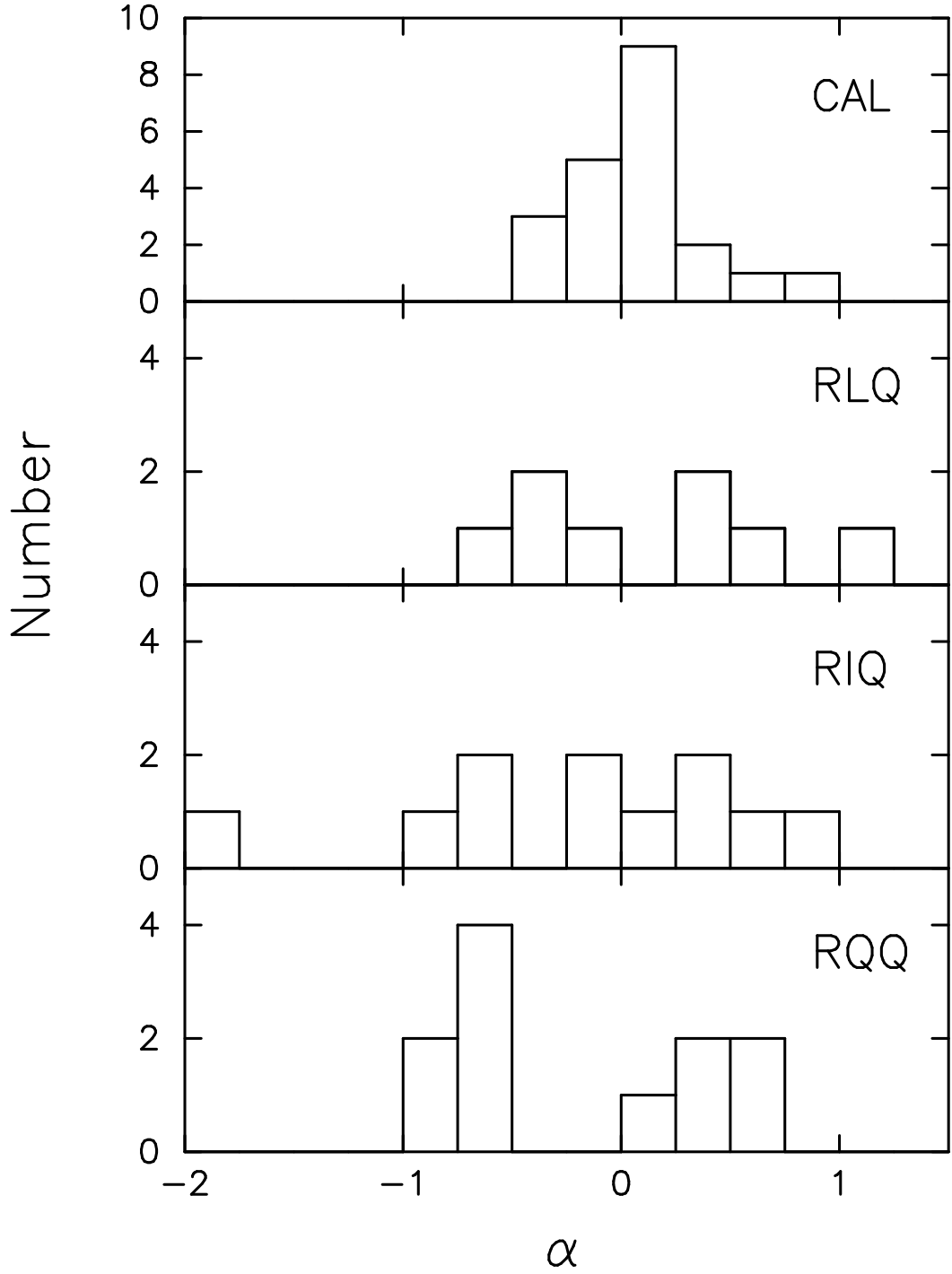} 
\caption{Comparison of radio spectral index histograms for the four 
source classes. 
}

\end{figure*}

\end{document}